\documentclass[twocolumn]{aastex61}

\received{}
\revised{}
\accepted{}
\submitjournal{ApJ}
\shorttitle{further evidence for a DPLH spectrum in GRB afterglows}
\shortauthors{Zhang et al.}
\defcitealias{Zhang15}{Paper~I}
\defcitealias{mang07}{M07}
\defcitealias{Resmi08}{RB08}

\begin{document}
\title{Modeling the Multi-band Afterglows of GRB~060614 and GRB~060908: Further Evidence for a Double Power-Law Hard Electron Energy Spectrum}

\correspondingauthor{Q. Zhang}
\email{zhangqiang@ihep.ac.cn}

\author{Q. Zhang}
\affiliation{Key Laboratory of Particle Astrophysics, Institute of High Energy Physics, Chinese Academy of Sciences, Beijing 100049, China}

\author{S. L. Xiong}
\affiliation{Key Laboratory of Particle Astrophysics, Institute of High Energy Physics, Chinese Academy of Sciences, Beijing 100049, China}

\author{L. M. Song}
\affiliation{Key Laboratory of Particle Astrophysics, Institute of High Energy Physics, Chinese Academy of Sciences, Beijing 100049, China}

\begin{abstract}
Electrons accelerated in relativistic collisionless shocks are usually assumed to follow a power-law energy distribution with an index of $p$.
Observationally, although most gamma-ray bursts (GRBs) have afterglows that are consistent with $p>2$, there are still a few GRBs suggestive of a hard ($p<2$) electron energy spectrum.
Our previous work showed that GRB~091127 gave strong evidence for a double power-law hard electron energy (DPLH) spectrum with $1<p_1<2$, $p_2>2$ and an ``injection break'' assumed as $\gamma_{\rm b}\propto \gamma^q$  in the highly relativistic regime, where $\gamma$ is the bulk Lorentz factor of the jet. In this paper, we show that GRB~060614 and GRB~060908 provide further evidence for such a DPLH spectrum. We interpret the multi-band afterglow of GRB~060614 with the DPLH model in an homogeneous interstellar medium by taking into account a continuous energy injection process, while for GRB~060908, a wind-like circumburst density profile is used. The two bursts, along with GRB~091127, suggest a similar behavior in the evolution of the injection break, with $q\sim0.5$. Whether this represents
a universal law of the injection break remains uncertain and more such afterglow observations are needed to test this conjecture.
\end{abstract}


\section{INTRODUCTION}\label{intro}
Gamma-Ray bursts (GRBs) are the most energetic stellar explosions in the universe. They produce a short prompt $\gamma$-ray emission followed by
a long-lived afterglow phase. The afterglows of GRBs are believed to originate from the synchrotron emission of shock-accelerated electrons produced by the interaction between the outflow and the external medium \citep{Rees92,Mes93,Mes97,Sari98,Cheva2000}. Particle acceleration is usually attributed to the \textit{Fermi} process \citep{Fermi54}, which results in a power-law (PL) energy distribution $N\left(E\right){\rm{d}}E \propto E^{-p}{\rm{d}}E$, with a cutoff at high energies. Some analytical and numerical studies indicate a nearly universal spectral index of $p\sim 2.2-2.4$ \citep[e.g.,][]{Bed98,Kirk00,Acht01,Lem03,Spit08}, though other studies suggest that there is a large range of possible values for $p$ of $1.5-4$ \citep{Bar04}. The values of $p$ derived from the spectral analysis of the multi-band afterglow \citep[e.g.][]{Cheva2000,Panai02,Starl08,Curran09,fong15,Li15,Wang15} or the X-ray data alone \citep[e.g.,][]{Shen06,Curran10} show a rather wide distribution, but most of them are consistent with $p>2$. Only a few GRBs, e.g., GRB~060908 \citep{Covino10}, GRB~091127 \citep{Filgas11,Troja12},
GRB~110918A \citep{Fred13} and GRB~140515A \citep{Mela15}, show very flat spectra in the optical band and require a hard ($p<2$) electron energy spectrum.

To explain those afterglows that cannot be well modeled with a standard ($p\gtrsim 2$) electron energy spectrum, two types of electron energy distributions were proposed in literature:
(1) a single PL electron energy distribution ($1<p<2$) with an exponential cutoff at a maximum electron Lorentz factor $\gamma_{\rm{M}}$ \citep{B01,Dai2001}; (2) a double PL electron energy distribution ($1<p_1<2$ and $p_2>2$) with an ``injection break'' $\gamma_{\rm{b}}$ \citep{Panai01,B04,Resmi08,Wang12}. A direct method to distinguish the two models is to see the passage of the injection break frequency $\nu_{\rm b}$ (i.e., the synchrotron frequency corresponding to $\gamma_{\rm b}$) through a certain band, e.g, from the optical to the near-infrared (NIR) bands. Our previous work \citep[][Paper I hereafter]{Zhang15} showed that GRB~091127 was such a case and gave strong evidence for the double PL hard electron spectrum model \citepalias[the so-called ``DPLH model'' in][]{Zhang15}.
The physical origin of $\gamma_{\rm b}$ is not clear. The DPLH model assumes $\gamma_{\rm b}\propto \gamma^q$ in the highly relativistic regime, here $\gamma$ is the bulk Lorentz factor of the jet.  \citetalias{Zhang15} found $q\sim0.6$ by modeling the multi-band afterglow of GRB~091127. Does this imply a universal evolution of the injection break? More GRB~091127-like bursts are needed to test this conjecture.

The ``smoking-gun'' evidence for a DPLH spectrum requires high-quality and multi-wavelength afterglow observations to provide detailed spectral information, in order to identify the existence of $\gamma_{\rm b}$ and its evolution behavior. In this paper, we
show that the multi-band afterglows of GRB~060614 and GRB~060908 can be well modeled by the DPLH model, thus providing further evidence for such a DPLH spectrum. Moreover, the two bursts, along with GRB~091127, seem to show a similar behavior in the evolution of the injection break.

Our paper is organized as follows. In Section \ref{obs}, we summarize the observational results of GRB~060614 and GRB~060908. Based on the work of \citet[][RB08 hereafter]{Resmi08}, the DPLH model for both a homogeneous interstellar medium (ISM) and a wind-like circumburst environment is described in Section \ref{model}. In this section we also extend the original model by taking into account a continuous energy injection process \citep{Sari00,Zhang01} to explain the afterglow of GRB~060614. In Section \ref{fitting}, we constrain the model parameters and then compare our model with the multi-band afterglow data. Finally, we present our conclusion and make some discussions in Section \ref{conclu}. The convention $F_{\nu}\propto\nu^{-\beta}t^{-\alpha}$ is adopted throughout the paper, where $\beta$ is the spectral index and $\alpha$ is the temporal decay index.
We use the standard notation $Q_x=Q/10^{x}$ with $Q$ being a generic quantity in cgs units and assume a concordance cosmology with $ H_0=70 ~\rm km \ s^{-1} Mpc^{-1}$, $\rm \Omega_M = 0.27$ and $\Omega_\Lambda =0.73$ \citep{jaro11}. All the quoted errors are given at a $1\sigma$ confidence level (CL) unless stated otherwise.

\section{OBSERVATIONAL RESULTS}\label{obs}
\subsection{GRB~060614} \label{obs1}
GRB~060614 triggered the {\it Swift} Burst Alert Telescope \citep[BAT;][]{Bart05} on 2006 June 14 at $T_0=$12:43:48 UT \citep{pars06} and  was also detected by {\it Konus-Wind} \citep{gole06}. The light curve (LC) shows an initial hard, bright peak lasting $\sim 5$~s followed by a long, somewhat softer extended emission, with a total duration of $T_{90}(15-350~\rm {keV})=102\pm3$~s \citep{bart06}. The spectrum of the initial pulse can be fitted in the 20~keV--2~MeV energy range by a PL with an exponential cutoff model, with the peak energy $E_{\rm pk}\sim302$~keV, while the spectrum of the remaining part of the burst can be described by a simple PL with photon index $2.13\pm0.03$ \citep{gole06}. The total fluence in the 20~keV--2~MeV energy range is $\sim4.1\times10^{-5}$ ${\rm erg\ cm^{-2}}$, of which the initial intense pulse contributes a fraction of $\sim20\%$ \citep{gole06}. With a redshift of $z=0.125$ \citep{fuga06,price06}, the isotropic equivalent energy was estimated as $E_{\gamma, \rm iso}=(2.5\pm0.4)\times10^{51}$ erg in the $1-10^4$~keV rest-frame energy band \citep[][M07 hereafter]{mang07}. In addition, GRB~060614 has null spectral lags, being consistent with typical short GRBs \citep{gehre06}.

The X-ray Telescope \citep[XRT;][]{Burr05} began observing the field 91~s after the BAT trigger \citep{pars06}. The X-ray afterglow of GRB~060614 exhibits a canonical LC which has been commonly observed in the {\it Swift} era \citep[e.g.,][]{Nou06,Zhang06,Evans09}. It begins with an initial
fast exponential decay, followed by a plateau with slope $\alpha_{\rm X,1}=0.11\pm0.03$; at $T_{\rm X, b1}=36.6\pm1.5$~ks, it steepens to a standard afterglow evolution with slope $\alpha_{\rm X,2}=1.03\pm0.01$; later on, the LC shows a further steepening to a slope $\alpha_{\rm X,3}=2.13\pm0.04$ at $T_{\rm X, b2}=104\pm13$~ks \citepalias[][see Figure \ref{LCs}]{mang07}. The X-ray data observed in the photon counting (PC) mode show no significant spectral evolution, with the spectral index $\beta_{\rm X}\sim 0.8$ \citepalias{mang07}.

The {\it Swift} Ultra-Violet/Optical Telescope \citep[UVOT;][]{Rom05} commenced observations 101~s after the BAT trigger \citep{holl06}. Besides, the R-band afterglow was detected by several ground telescopes \citep[e.g.,][]{dell06,fren06,fynb06,gal06}. \citetalias{mang07} presented detailed spectral and temporal analysis of the optical/ultraviolet (UV) afterglow, below we summarize their main results. The optical/UV LCs show achromatic breaks with the X-ray afterglow, i.e., $t_{\rm UVO,b1}=29.7\pm 2.7$~ks and $t_{\rm UVO,b2}=117.2\pm 2.7$~ks. The decay slopes after the two breaks are
$\alpha_{\rm UVO,2}=1.11\pm0.03$ and $\alpha_{\rm UVO,3}=2.44\pm0.05$, respectively. On the whole, the X-ray/UV/optical LCs have marginally consistent evolutions after $\sim 30$~ks. What is puzzling is that the initial slope $\alpha_{\rm UVO,1}$ is dependent on wavelength: the UV LCs show nearly flat evolutions while the optical LCs rise slowly with slopes from $\sim(-0.38)$ to $\sim(-0.17)$ (see Figure \ref{LCs}).
The spectral energy distributions (SEDs) of the afterglow from optical to X-rays show a spectral break passing through the optical/UV band between $\sim 10$ and $\sim30$~ks. The break frequency at 10~ks is around $1.0\times10^{15}$~Hz \citepalias[see Figure 7 of][]{mang07}. At this time, the optical/UV and X-ray afterglows have spectral indices $\beta_{\rm UVO}=0.30\pm0.09$ and $\beta_{\rm X}=0.84\pm0.04$,
respectively. At later times ($t\gtrsim30$~ks), the spectral index in the optical/UV band changes to be consistent with that of X-rays.
Fits of the broad-band SEDs imply a weak host extinction $A_{V, \rm h}=0.05\pm0.01$ \citepalias{mang07}.

In addition, deep optical/NIR follow-ups of GRB~060614 show no evidence for an associated supernova down to very strict limits; the GRB host is a
very faint star-forming galaxy with a specific star formation rate lower than most long GRB hosts; the GRB counterpart resides in the outskirts
of the host \citep{dell06,fynb06,gal06}. The recent discovery of a distinct NIR excess at about 13.6 days after the burst suggests
a possible kilonova (or macronova) origin \citep{jin15,yang15}. Together with the vanishing time lags of the prompt emission, all these point towards
a different origin from typical long GRBs; it is likely to be of a subclass of merger-type short GRBs \citep{gehre06,zhang07}.

\subsection{GRB~060908}\label{obs2}
GRB~060908 triggered the {\it Swift}/BAT on 2006 September 14 at $T_{\rm BAT}=$08:57:22.34 UT \citep{evans06}. Further analysis found the onset of the GRB occurs 12.96~s before the trigger time, i.e., $T_0=T_{\rm BAT}-12.96~{\rm s}$ \citep{Covino10}. So the time used in this work is relative to $T_0$.
The BAT LC shows a multi-peaked structure with a total duration of $T_{90}(15-350~\rm {keV})=19.3\pm0.2$~s \citep{palmer06}.
The time-averaged spectrum is best fit by a simple PL and can be alternatively fit by a Band function \citep{band93} with the high-energy photon index fixed. With a redshift of $z=1.884$ \citep{fynbo09}, the latter spectral model gives the rest-frame peak energy $E_{\rm p,i}\sim380$~keV and the isotropic equivalent energy $E_{\gamma, \rm iso}=(6.2\pm0.4)\times10^{52}$~erg in the rest-frame $1-10^4$~keV energy band \citep{Covino10}.

The XRT began observing the field 72~s after the BAT trigger\citep{evans06}. The spectra were modeled with an absorbed power-law, which gave the spectral index $\beta_{\rm X}=1.17^{+0.25}_{-0.22}$ and the host absorbing column density $N_{\rm{H}}\sim 8.3\times10^{21}$ cm$^{-2}$. The LC is characterised by a constant PL decay with index $\alpha_{\rm X}=1.12^{+0.05}_{-0.02}$, while from $\sim 200$ to $\sim 1000$~s a complex flaring activity is superposed on the underlying decay \citep[][see Figure \ref{LCs2}]{Covino10}.

The UVOT commenced observations 80~s after the BAT trigger \citep{morgan06}. The optical/NIR afterglow was also monitored by several ground-based telescopes \citep[e.g.,][]{andr06,anto06,nyse06,wier06}.
The LCs can be described by a broken PL with the initial decay index $\alpha_{\rm optNIR,1}=1.48\pm0.25$, the break time $t_{\rm optNIR,1}=138^{+167}_{-43}$~s and the post-break decay index $\alpha_{\rm optNIR,2}=1.05\pm0.03$ \citep{Covino10}. There seems to be another
break at $\sim10^3-10^4$~s, with the post-break decay slope of $1.1-1.4$ (see Figure \ref{LCs2}). However, this break time cannot be well constrained by the data \citep{Covino10}. In Subsection \ref{fitting2}, we will show that such a late break is actually required by the afterglow modeling.
The spectral analysis at 800 and 8000~s shows rather flat spectra with index $\beta_{\rm optNIR}=0.33^{+0.25}_{-0.29}$ and host dust extinction
$E(B-V)\sim0.03$ \citep{Covino10}.

\section{MODEL}\label{model}
Several clues should be considered before establishing the afterglow model for both GRBs. For GRB~060614: (i) The two achromatic breaks ($t_{\rm b,1}\equiv t_{\rm UVO,b1}\approx t_{\rm X,b1}$ and $t_{\rm b,2}\equiv t_{\rm UVO,b2}\approx t_{\rm X,b2}$) shown in the multi-band LCs require a hydrodynamical origin. This canonical afterglow behavior was well described in \citet{Zhang06}. The first break is possibly an ``energy-injection break'', implying the end of a continuous
energy injection into the forward shock \citep{Sari00,Zhang01}, while the second break is most likely the so-called  ``jet break'' \citep{Rhoads99,Sari99}; (ii) The early flat spectrum ($\beta_{\rm UVO}\sim 0.3$) in the optical/UV band definitely requires a hard electron energy spectrum; (iii) There should be a spectral break between the optical/UV and the X-ray bands, but neither the minimum synchrotron frequency $\nu_{\rm m}$ nor the cooling frequency $\nu_{\rm c}$ can accommodate the observations \footnote{The reasons are as follows: (i) Although the passage of $\nu_{\rm m}$ can produce a spectral evolution and  slow-rising optical LCs \citepalias{mang07}, this requires $\nu_{\rm opt}<\nu_{\rm m}$. The model predicted spectral index in this regime is $\beta_{\rm opt}=-1/3$ which is inconsistent with the observed value ($\sim0.3$); (ii) If the observed break frequency is $\nu_{\rm c}$, it suggests a hard electron energy spectrum with $p=2\beta_{\rm UVO}+1\sim1.6$. However, the single PL hard electron spectrum model of \citet{Dai2001} predicts the post-jet-break decay slope should be $\sim 1.9$ which is substantially lower than the observed value \citepalias[$\sim2.4$;][]{mang07}.}. For GRB~060908, the SED analysis also requires a hard electron energy distribution and some kind of spectral break between the optical/NIR and the X-ray bands. The single PL hard electron spectrum model of \citet{Dai2001} with $\nu_{\rm optNIR}<\nu_{\rm c}<\nu_{\rm X}$  has difficulties in explaining the observations, since the model
predicts $\beta_{\rm X}\sim 0.8$ which is obviously lower than the observed value; the predicted decay slopes are also inconsistent with the observations. Therefore, the DPLH model is a natural choice.

Using the derived spectral and temporal indices of \citetalias{Resmi08} (their Table 2), we found the afterglow properties of GRB~060614 can be well reproduced by the DPLH model for an ISM medium when an additional energy injection is invoked, while the properties of GRB~060908 can be well explained when a wind-like circumburst density profile is used. In this section, we give a basic description of the DPLH model and present relevant formulas which will be used in Section \ref{fitting}. We refer the reader to \citetalias{Resmi08} for more details.

The DPLH spectrum with indices $1<p_1<2$ and $p_2>2$ is represented as \citepalias{Resmi08}
\begin{equation}
N\left(\gamma_{\rm{e}}\right)= C_{\rm{e}}\left\{
\begin{array}{ll}
\left(\frac{\gamma_{\rm{e}}}{\gamma_{\rm{b}}}\right)^{-p_1}, &\gamma_{\rm{m}}\leqslant\gamma_{\rm{e}}<\gamma_{\rm{b}},  \\
\left(\frac{\gamma_{\rm{e}}}{\gamma_{\rm{b}}}\right)^{-p_2}, &\gamma_{\rm{e}}\geqslant\gamma_{\rm{b}},  \\
\end{array}
\right.
\end{equation}
where $C_{\rm{e}}$ is the normalization constant, $\gamma_{\rm{m}}$ is minimum electron Lorentz factors, and $\gamma_{\rm{b}}$ is the injection break. The physical origin of $\gamma_{\rm{b}}$ is not clear,
\citetalias{Resmi08} assumed that it is a function of $\beta\gamma$ to accommodate the non-relativistic regime of expansion, i.e.,
\begin{equation}
\gamma_{\rm{b}}=\xi \left(\beta \gamma\right)^q,
\end{equation}
where $\xi$ is a constant of proportionality, $\beta=\sqrt{1-\gamma^{-2}}$ is the dimensionless bulk velocity, and $q$ is assumed to be a constant for simplicity.

For a a relativistic shock propagating through a cold medium with particle density $n$, the post-shock
particle density and energy density are $4\gamma n$ and $4\gamma(\gamma-1)n m_{\rm p} c^2$, respectively \citep{Sari98}, from which one derives the minimum Lorentz factor \citepalias{Resmi08}
\begin{equation}
\gamma_{\rm{m}}=\left(f_{\rm{p}} \frac{m_{\rm{p}}}{m_{\rm{e}}}\frac{\epsilon_{\rm{e}}}{\xi^{2-p_1}} \right)^{\frac{1}{p_1-1}} \beta^{-\frac{q\left(2-p_1\right)}{p_1-1}} \left(\gamma-1\right)^{\frac{1}{p_1-1}} \gamma^{-\frac{q\left(2-p_1\right)}{p_1-1}},
\end{equation}
where $m_{\rm p}$ and $m_{\rm e}$ are the proton and electron rest mass, respectively; $\epsilon_{\rm e}$ is the fraction of
shock energy carried by electrons, and $f_{\rm{p}}=[(2-p_1)(p_2-2)]/[(p_1-1)(p_2-p_1)]$.

We calculate the break frequencies of synchrotron spectra $\nu_{\rm{m}}$, $\nu_{\rm{b}}$, $\nu_{\rm{c}}$  and the peak flux $F_{\nu,\rm{max}}$ according to the formulas given by \citet{Wijers99}:
\begin{eqnarray}\label{vm1}
\nu_{\rm{m}} &=& \frac{x_p}{1+z}\frac{q_{\rm{e}}B'}{\pi m_{\rm{e}}c} \gamma \gamma_{\rm{m}}^2,\\ \label{vbc}
\nu_{\rm{b,c}} &=& \frac{0.286}{1+z}\frac{q_{\rm{e}}B'}{\pi m_{\rm{e}}c} \gamma \gamma_{\rm{b,c}}^2,\\ \label{fmax}
F_{\nu,\rm{max}} &=& \frac{\sqrt{3}\phi_{p} N_{\rm e} q_{\rm{e}}^3 \left(1+z\right)}{4\pi d_{\rm{L}}^2 m_{\rm{e}} c^2 } B'\gamma,
\end{eqnarray}
where $N_{\rm e}$ is the total number of swept-up electrons, $q_{\rm{e}}$ is the electron charge, $B'=\left(32\pi n m_{\rm{p}} c^2 \epsilon_{\rm{B}}\right)^{1/2}\gamma$ is the post-shock magnetic field density, $\epsilon_{\rm B}$ is the fraction of
shock energy carried by magnetic fields, $d_{\rm{L}}$ is the luminosity distance corresponding to the redshift $z$, $\gamma_{\rm{c}}=6\pi m_{\rm{e}}c/\left(\sigma_{\rm{T}}\gamma B'^2 t\right)$ is the cooling Lorentz factor of electrons. $x_{p}$ and $\phi_{p}$ represent the dimensionless peak frequency and the peak flux, respectively. Their dependence on $p$ can be obtained from \citet{Wijers99}.

For the adiabatic self-similar evolution of a spherical blastwave, the radius $r$ and bulk Lorentz factor $\gamma$ evolve as
$r=[(17-4k)(4-k)Et/4\pi A m_{\rm p} c (1+z)]^{1/\left(4-k\right)}$ and $\gamma=[(17-4k)E(1+z)^{3-k}/4^{5-k}(4-k)^{3-k}\pi A m_{\rm p}c^{5-k}t^{3-k}]^{1/2\left(4-k\right)}$ in the ultra-relativistic regime \citep{BM76,Sari98,Cheva2000,Gao13}.
The above derivation used the density profile $n=A r^{-k}$, $k=0$ for ISM and $k=2$ for wind medium. By substituting these expressions in Equations (\ref{vm1})--(\ref{fmax}), one derives \footnote{Different from \citetalias{Resmi08}, we did not consider the effect of sideways expansion
in the derivation of Equations (\ref{vm2})-(\ref{fmax2w}). This can be seen as a reasonable approximate in the ultra-relativistic regime as long as
the inverse Lorentz factor has not exceeded the initial jet opening angle \citep{Rhoads99}. The coefficients in these equations are consistent
with those of \citetalias{Resmi08} within a factor of a few that may be due to minor differences in the treatment of dynamics.}
\begin{eqnarray} \label{vm2}
\nu_{\rm{m}} &=& 8.2\times 10^{6}\left(1833 f_{\rm{p}}\right)^{\frac{2}{p_1-1}} \left(37.2\right)^{\frac{1-q\left(2-p_1\right)}{p_1-1}} \frac{x_{p_1}}{1+z} \nonumber\\
             &  & \xi^{\frac{-2\left(2-p_1\right)}{p_1-1}}\epsilon_{\rm{e}}^{\frac{2}{p_1-1}} \epsilon_{\rm{B,-2}}^{1/2}
             E_{52}^{\frac{p_1-q\left(2-p_1\right)}{4\left(p_1-1\right)}}n_{0}^{\frac{p_1-2+q\left(2-p_1\right)}{4}}     \nonumber\\
             &  & \left(\frac{t_{\rm{d}}}{1+z}\right)^{\frac{-3\left[p_1-q\left(2-p_1\right)\right]}{4\left(p_1-1\right)}} \rm{Hz},  \\\label{vc2}
\nu_{\rm{c}} &=& 1.5\times 10^{15} \epsilon_{\rm{B,-2}}^{-3/2} E_{52}^{-1/2} n_0^{-1} \left[t_{\rm{d}}(1+z)\right]^{-1/2} \rm{Hz}, \\ \label{vb}
\nu_{\rm{b}} &=& 3.8\times 10^{5} \frac{\left(6.1\right)^{1+2q}}{1+z} \xi^2 \epsilon_{\rm{B,-2}}^{1/2} E_{52}^{\frac{1+q}{4}}
                 n_{0}^{\frac{1-q}{4}} \nonumber\\
             &  &\left(\frac{t_{\rm{d}}}{1+z}\right)^{-\frac{3\left(1+q\right)}{4}} \rm{Hz}, \\ \label{fmax2}
F_{\nu,\rm{max}} &=& 6.8\times 10^3 \phi_{p_1} \epsilon_{\rm{B,-2}}^{1/2} E_{52} n_0^{1/2} d_{\rm{L,28}}^{-2} \left(1+z\right) \rm{\mu Jy},
\end{eqnarray}
for the ISM case, and
\begin{eqnarray}\label{vm2w}
\nu_{\rm m} &=& 5.8\times10^6(13.8)^y(183.3f_{\rm p})^{\frac{2}{p_1-1}}\frac{x_{p_1}}{1+z}\xi^{\frac{-2\left(2-p_1\right)}{p_1-1}} \nonumber \\
            & & \epsilon_{\rm e,-1}^{\frac{2}{p_1-1}}\epsilon_{\rm B,-2}^{1/2}
            E_{52}^{y/2}A_{\ast}^{\frac{1-y}{2}}\left(\frac{t_{\rm d}}{1+z}\right)^{-\frac{2+y}{2}}{\rm Hz}, \\ \label{vc2w}
\nu_{\rm{c}} &=& \frac{1.6\times 10^{15}}{(1+z)^3} \epsilon_{\rm{B,-2}}^{-3/2} E_{52}^{1/2} A_{\ast}^{-2}
\left(\frac{t_{\rm d}}{1+z}\right)^{1/2} {\rm Hz}, \\ \label{vbw}
\nu_{\rm{b}} &=& 1.6\times 10^{6} \frac{\left(13.8\right)^{q}}{1+z} \xi^2 \epsilon_{\rm{B,-2}}^{1/2} E_{52}^{q/2}A_{\ast}^{\frac{1-q}{2}}\nonumber \\
             & & \left(\frac{t_{\rm{d}}}{1+z}\right)^{-\frac{2+q}{2}} {\rm Hz}, \\ \label{fmax2w}
F_{\nu,\rm{max}} &=& 20.6 \phi_{p_1}(1+z) \epsilon_{\rm{B,-2}}^{1/2} E_{52}^{1/2} A_{\ast}d_{\rm{L,28}}^{-2} \nonumber \\
                 & & \left(\frac{t_{\rm d}}{1+z}\right)^{-1/2} {\rm mJy},
\end{eqnarray}
for the wind case\footnote{We note the exponent of $\xi$ in Equation (\ref{vm2w}) and the exponent of $A_{\ast}$ in Equation (\ref{vbw}) are different from the results of \citetalias{Resmi08} (see their Equations (13) and (15)). Their expression of $\nu_{\rm b}$ also missed out a factor of $1/(1+z)$. We have carefully checked our derivations to make sure that our results are robust. Here $t_{\rm{d}}$ is the time in days, $A_{\ast}$ and $A$ are related by $A=3\times10^{35}A_{\ast}$~cm$^{-1}$ and $y=\left[1-q\left(2-p_1\right)\right]/\left(p_1-1\right)$.}

The evolution of the synchrotron flux density at a given frequency ($F_\nu$) relies on the order of the three break frequencies and the regime
in which $\nu$ resides. Below we give only some scaling laws\footnote{We refer the reader to \citetalias{Resmi08} for a complete reference of the scaling relationships for the spectral breaks and $F_{\nu}$ in various spectral regimes (but without an energy injection).} for $F_{\nu}$ and $\nu_{\rm b}$ that will be used in Section \ref{fitting}. Since the synchrotron self-absorption process is not relevant, we do not consider it in this work.

For GRB~060908, according to Equations (\ref{vm2w})-(\ref{fmax2w}), the relevant spectral regimes and flux densities are:

(i) $\nu_{\rm m}<\nu<\min{(\nu_{\rm b}, \nu_{\rm c})}$,
\begin{equation}
F_{\nu}=F_{\nu,\rm{max}}\left(\frac{\nu}{\nu_{\rm{m}}}\right)^{-\frac{p_1-1}{2}}\propto
t^{\frac{1}{4}\left(2q-p_1q-2p_1-1\right)}. \label{fv1w}
\end{equation}
(ii) $\nu_{\rm m}<\nu_{\rm b}<\nu<\nu_{\rm c}$,
\begin{eqnarray}
F_{\nu}&=&F_{\nu,\rm{max}}\left(\frac{\nu_{\rm{b}}}{\nu_{\rm{m}}}\right)^{-\frac{p_1-1}{2}}
\left(\frac{\nu}{\nu_{\rm{b}}}\right)^{-\frac{p_2-1}{2}} \nonumber\\
    &\propto& t^{\frac{1}{4}\left(2q-p_2q-2p_2-1\right)}. \label{fv2w}
\end{eqnarray}
(iii) $\nu>\max{(\nu_{\rm b}, \nu_{\rm c})}>\nu_{\rm m}$,
\begin{eqnarray}
F_{\nu} &=& F_{\nu,\rm max}\left(\frac{\nu_{\rm c}}{\nu_{\rm m}}\right)^{-\frac{p_1-1}{2}}\left(\frac{\nu_{\rm b}}{\nu_{\rm c}}\right)^{-\frac{p_1}{2}}\left(\frac{\nu}{\nu_{\rm b}}\right)^{-\frac{p_2}{2}} \nonumber \\
        &=& F_{\nu,\rm max}\left(\frac{\nu_{\rm b}}{\nu_{\rm m}}\right)^{-\frac{p_1-1}{2}}\left(\frac{\nu_{\rm c}}{\nu_{\rm b}}\right)^{-\frac{p_2-1}{2}}\left(\frac{\nu}{\nu_{\rm c}}\right)^{-\frac{p_2}{2}} \nonumber \\
        &\propto&  t^{\frac{1}{4}\left(2q-p_2q-2p_2\right)}. \label{fv3w}
\end{eqnarray}

For GRB~060614, the situation is somewhat more complicated. Besides the adiabatic self-similar evolution phase, these should be a continuous
energy injection process before $\sim30$~ks and a jet break at about 117~ks.
The injected energy can be provided by a long-lived central engine \citep{Dai98,Zhang01} or by slower material with significant energy which gradually piles up onto the decelerating ejecta and ``refreshes'' it \citep{Rees98,Sari00}. Here we do not consider a specific energy injection mechanism and generally assume that the isotropic equivalent blastwave energy $E$ evolves as
\begin{equation}\label{Et}
E\left(t\right)=\left\{
\begin{array}{ll}
E_{\rm f}\left(\frac{t}{t_{\rm f}}\right)^{1-e}, & t_{\rm i}\leqslant t<t_{\rm f},\\
E_{\rm f}, & t\geqslant t_{\rm f},
\end{array}
\right.
\end{equation}
where $t_{\rm i}$ is the time when the assumed PL energy injection ($E\propto t^{1-e})$ begins,
$t_{\rm f}$ is the end time of the energy injection, $E_{\rm f}$ is the final blastwave energy, and $e<1$ is required for an effective energy injection. When this energy injection is taken into account, the blastwave energy $E$ in Equations (\ref{vm2})-(\ref{fmax2}) should be replaced with Equation (\ref{Et}).

According to Equations (\ref{vm2})-(\ref{fmax2}), the relevant spectral regimes and flux densities are:

(i) $\nu_{\rm{m}}<\nu<\nu_{\rm{b}}<\nu_{\rm{c}}$,
\begin{equation}
F_{\nu}=F_{\nu,\rm{max}}\left(\frac{\nu}{\nu_{\rm{m}}}\right)^{-\frac{p_1-1}{2}}\propto t^{\left[\left(1-e\right)-\frac{\left(2+e\right)\left(p_1+p_1 q-2q\right)}{8}\right]}. \label{fv1}
\end{equation}
(ii) $\nu_{\rm{m}}<\nu_{\rm{b}}<\nu<\nu_{\rm{c}}$,
\begin{eqnarray}
F_{\nu}&=&F_{\nu,\rm{max}}\left(\frac{\nu_{\rm{b}}}{\nu_{\rm{m}}}\right)^{-\frac{p_1-1}{2}}
\left(\frac{\nu}{\nu_{\rm{b}}}\right)^{-\frac{p_2-1}{2}} \nonumber\\
    &\propto& t^{\left[\left(1-e\right)-\frac{\left(2+e\right)\left(p_2+p_2 q-2q\right)}{8}\right]}. \label{fv2}
\end{eqnarray}

According to Equation (\ref{vb}), the injection break frequency $\nu_{\rm b}$ scales as
\begin{equation}
\nu_{\rm b}\propto t^{-\frac{\left(2+e\right)\left(1+q\right)}{4}}.
\label{vb_scale}
\end{equation}

After the end of the energy injection ($t\geqslant t_{\rm f}$), the blastwave enters an adiabatic evolution phase and the corresponding scaling relationships can be easily obtained by setting $e=1$ in Equations (\ref{fv1})--(\ref{vb_scale}).

We next discuss the physical origin of the jet break of GRB~060614.
For a simplified conical jet with a half-opening angle $\theta_{\rm j}$, as it decelerates, the radiation beaming angle ($1/\gamma$) would eventually exceed the jet half-opening angle, i.e., $1/\gamma>\theta_{\rm j}$. At this time, a jet break may occur in the afterglow LC. Two effects could result in a jet break: the first is the pure jet-edge effect which steepens the LC by $t^{-3/4}$
for an ISM medium \citep{Mes99}; the second effect is  caused by sideways expansion, which has important effects on the hydrodynamics when $1/\gamma\gtrsim\theta_{\rm{j}}$ is satisfied and the post-jet-break flux decays as $t^{-p}$ for a normal electron energy spectrum with index $p>2$ \citep{Rhoads99,Sari99}.

For GRB~060614, the jet break should be a result of significant sideways expansion rather than the jet-edge effect. The reasons are as follows: (i) The post-jet-break decay (in the optical/UV band) caused by the edge effect would have a slope $\sim1.1+0.75=1.85$, which is substantially lower than the observed value ($\sim 2.44$); (ii) Using the expression (their Equation (11)) given by \citet{Wang12} who considered the effect of sideways expansion in a similar DPLH model and the obtained parameter values ($p_2$ and $q$) in Subsection \ref{fitting1}, we estimate the post-jet-break slope to be $\sim2.48$ which is excellently consistent with the observed value.

Based on the work of \citet{Wang12}, we give the scaling law for $F_{\rm \nu}$ in the post-jet-break phase straightforwardly.
For $\nu_{\rm{m}}<\nu_{\rm{b}}<\nu<\nu_{\rm{c}}$,
\begin{equation}
F_{\nu}\propto t^{-\frac{q\left(p_2-2\right)+\left(p_2+2\right)}{2}},~~~t>t_{\rm j},
\label{fv_post}
\end{equation}
where $t_{\rm j}$ is the jet-break time.

\section{PARAMETER CONSTRAINT AND AFTERGLOW MODELING} \label{fitting}
\subsection{GRB~060614} \label{fitting1}
Before constraining the free parameters ($p_1$, $p_2$, $q$, $e$, $\epsilon_{\rm{e}}$, $\epsilon_{\rm{B}}$, $\xi$, $E_{\rm f}$ and $n$),
we first summarize the relevant observational results of GRB~060614: (i) $\beta_{\rm UVO}(10~{\rm ks})=0.30\pm0.09$,
$\beta_{\rm X}=0.84\pm0.04$; (ii) $\alpha_{\rm X,1}=0.11\pm0.03$; $\alpha_{\rm UVO,2}=1.11\pm0.03\approx \alpha_{\rm X,2}$;
(iii) $t_{\rm b,1}=29.7\pm2.7$~ks, $t_{\rm b,2}=117.2\pm2.7$~ks;
(iv) $\tilde{\nu}_{\rm b}(10~{\rm ks})\approx 1.0\times10^{15}$~Hz\footnote{Here and below we use $\tilde{\nu}_{\rm b}$ to denote the observed break frequency in the SEDs, in order to distinguish with the injection break frequency $\nu_{\rm b}$ in our model.}; (v) $\tilde{\nu}_{\rm b}(30~{\rm ks})\lesssim \nu_{\rm R}$, since the SED shows that
the break frequency has just crossed the R-band at about 30~ks; (vi) the initial decay slope of the $R$-band LC $\alpha_{\rm R,1}=-0.38\pm 0.14$; (vii) $\alpha_{\rm UVO,3}=2.44\pm0.05\approx \alpha_{\rm X,3}$.
In this section we use conditions (i)--(iv) to constrain the model parameters, and use (v)--(vii) for consistency checks.

Using condition (i), we get $p_1=2\beta_{\rm UVO}(10~{\rm ks})+1=1.60\pm0.18$ and $p_2=2\beta_{\rm X}+1=2.68\pm0.08$. The values of $q$ and $e$
can be obtained from condition (ii) and Equation (\ref{fv2}), i.e.,
\begin{eqnarray}
\frac{\left(2+e\right)\left(p_2+p_2 q-2q\right)}{8}-\left(1-e\right) &=& 0.11\pm0.03, \\
\frac{3\left(p_2+p_2q-2q\right)}{8} &=& 1.11\pm0.03. \label{q_eq}
\end{eqnarray}
Solving these equations gives $q=0.41\pm0.20$ and $e=0.27\pm0.04$. With these values, we test our model predictions with conditions
(v)--(vii). First, Equation (\ref{vb_scale}) gives $\nu_{\rm b}\propto t^{-0.80\pm0.12}$ during the energy injection phase, then, with condition (iv) we have $\nu_{\rm b}(30~{\rm ks})\approx 4.2\times10^{14}~{\rm Hz}$, which is excellently consistent with condition (v). Second, based on Equation (\ref{fv1}), the predicted initial $R$-band decay slope is $-0.32\pm 0.09$ that is consistent with the observational results (condition (vi)) within 1~$\sigma$ errors. Finally, we estimate the post-jet-break decay slope from Equation (\ref{fv_post}) and the obtained value is $2.48\pm0.09$,
which is in perfect accord with that of the optical/UV afterglow, and marginally consistent with that of the X-ray afterglow. These exciting results encourage us to have a further check of our model by modeling the afterglow LCs. In the following calculations, we adopt $p_1=1.6$, $p_2=2.68$, $q=0.41$ and $e=0.27$.

In the normal decay phase ($t_{\rm b,1}<t<t_{\rm b,2}$), we have $\nu_{\rm m}<\nu_{\rm b}<\nu_{\rm UVO}<\nu_{\rm X}<\nu_{\rm c}$. Following Equations (\ref{vm2})--(\ref{fmax2}) and (\ref{fv2}), one derives\footnote{$x_{p_1}=0.85$ and $\phi_{ p_1}=0.5$ were adopted in the derivations according to \citet{Wijers99} and our obtained $p_1=1.6$. The same values were used to calculate $\nu_{\rm m}$ and $F_{\nu,\rm max}$ for GRB~060908 in Subsection \ref{fitting2}.}
\begin{eqnarray}\label{vm3}
\nu_{\rm m}&=&1.1\times10^{10}\xi_4^{-1.33}\epsilon_{\rm e,-1}^{3.33}\epsilon_{\rm B,-2}^{1/2}E_{\rm f,52}^{0.6}n_0^{-0.06}t_{\rm d}^{-1.8}{\rm Hz}, \\
\nu_{\rm b}&=&1.0\times10^{15}\xi_4^2\epsilon_{\rm B,-2}^{1/2}E_{\rm f,52}^{0.35} n_0^{0.15} t_{\rm d}^{-1.06}~{\rm Hz}, \\
\nu_{\rm c}&=&1.4\times10^{15}\epsilon_{\rm B,-2}^{-3/2}E_{\rm f,52}^{-1/2}n_0^{-1}t_{\rm d}^{-1/2}~{\rm Hz}, \\ \label{fv3}
F_{\nu_{\rm R}}&=&7.5\times10^3\xi_4^{0.68}\epsilon_{\rm e,-1}\epsilon_{\rm B,-2}^{0.92}E_{\rm f,52}^{1.37}n_0^{0.56}t_{\rm d}^{-1.11}\mu{\rm Jy}.
\end{eqnarray}

To constrain the parameters, we require that (i) the $R$-band flux at 52~ks is $F_{\nu_{\rm R}}(52~{\rm ks})=55.9~\mu{\rm Jy}$\footnote{This value has been corrected for Galactic and host galaxy extinction with $A_{V,{\rm G}}=0.07$ and $A_{V,\rm h}=0.05$, respectively, according to the results of \citetalias{mang07}.}, (ii) $\nu_{\rm b}(10~{\rm ks})=1.0\times 10^{15}$~Hz, and (iii) $\nu_{\rm{c}}$ should well above 10~keV at the last measurement of the X-ray afterglow, i.e., $\nu_{\rm c}(2\times10^6~{\rm s})>10~{\rm keV}$. After a simple calculation, we get
\begin{eqnarray} \label{cond1}
\epsilon_{\rm B,-2}n_0^{2/3} &=& 1.95\times10^{-3}\epsilon_{\rm e,-1}^{-4/3}E_{\rm f,52}^{-5/3}, \\ \label{cond2}
\xi_4 &=& 1.5\epsilon_{\rm e,-1}^{-1/3}E_{\rm f,52}^{0.24}n_0^{0.09}, \\ \label{cond3}
\epsilon_{\rm e,-1} &>& 0.84E_{\rm f,52}^{-1}.
\end{eqnarray}
With only two equations, the model parameters ($\epsilon_{\rm e}$, $\epsilon_{\rm B}$, $\xi$, $E_{\rm f}$ and $n$) are strongly degenerate.
Here we adopt a typical value of  $\epsilon_{\rm e,-1}=1$, which has been supported by recent large sample afterglow modelings \citep[e.g.,][]{nava14,sant14,beni17}. $E_{\rm f}$ is the final blastwave energy after the energy injection, of which the mechanism was not specified above. Here we simply assume an equivalent prompt emission efficiency of $\eta_{\gamma}=E_\gamma/(E_\gamma+E_{\rm f})=10\%$ and leave the discussion
on the energy injection mechanism in Section \ref{conclu}. With $E_{\gamma}=2.5\times10^{51}$~erg and $\eta_\gamma=10\%$, we obtain $E_{\rm f,52}=2.25$ and Equation (\ref{cond3}) is naturally satisfied.  By substituting these values in Equation (\ref{cond1}), we get $\epsilon_{\rm{B},-2} n_{0}^{2/3}=5.0\times10^{-4}$.
The values of $\epsilon_{\rm B}$  and $n$ cannot be well constrained since both of them are highly uncertain parameters and vary over several
orders of magnitude. By modeling the multi-band afterglows of 38 short GRBs, \citet{fong15} gave a median density of $n\sim10^{-3}-10^{-2}$~cm$^{-3}$
, and found that 80\%--95\% of bursts have densities of $n\lesssim1$~cm$^{-3}$. For GRB 060614, if we take $n_0=10^{-1}$ to $10^{-3}$, we get
$\epsilon_{\rm B,-2}=2.3\times10^{-3}$ to $5\times10^{-2}$. These values are well consistent with the recent results of \citet{sant14} and \citet{Duran14}, who found the distribution of $\epsilon_{\rm{B}}$ has a range of $\sim 10^{-8} - 10^{-3}$ with a median value of $\sim$ a few $\times 10^{-5}$. In the following calculations, we adopt $n_0=0.01$ and $\epsilon_{\rm B,-2}=1.1\times10^{-2}$. Finally, we substitute the above
values in Equation (\ref{cond2}) and get $\xi_4=1.2$. We note $\xi$ is weakly dependent on other parameters and can be well constrained; it is around
$10^4$, varying within a factor of two.

Since we interpret the achromatic break at $t_{\rm b,2}$ as a jet break, we can estimate the half-opening angle of the jet according to $\theta_{\rm{j}}\sim \gamma(t_{\rm j})^{-1}$ \citep{Rhoads99,Sari99}. We thus have
\begin{equation}
\theta_{\rm{j}}=9.4\degr~ E_{52}^{-1/8} n_0^{1/8} \left(\frac{t_{\rm{b,2,d}}}{1+z}\right)^{3/8}= 5.1\degr.
\end{equation}
Using $\gamma(t_{\rm j})\sim \theta_{\rm j}^{-1}=11.2$ and $\gamma(t>t_{\rm j})\propto t^{-1/2}$\citep{Rhoads99}, we have
$\gamma(2\times10^6~{\rm s})\sim2.7$, which suggests a mildly relativistic jet even at the end of the X-ray observations. Therefore,
our explanation of the entire afterglow of GRB~060614 in the highly relativistic regime is self-consistent.

\begin{figure}[ht!]
\plotone{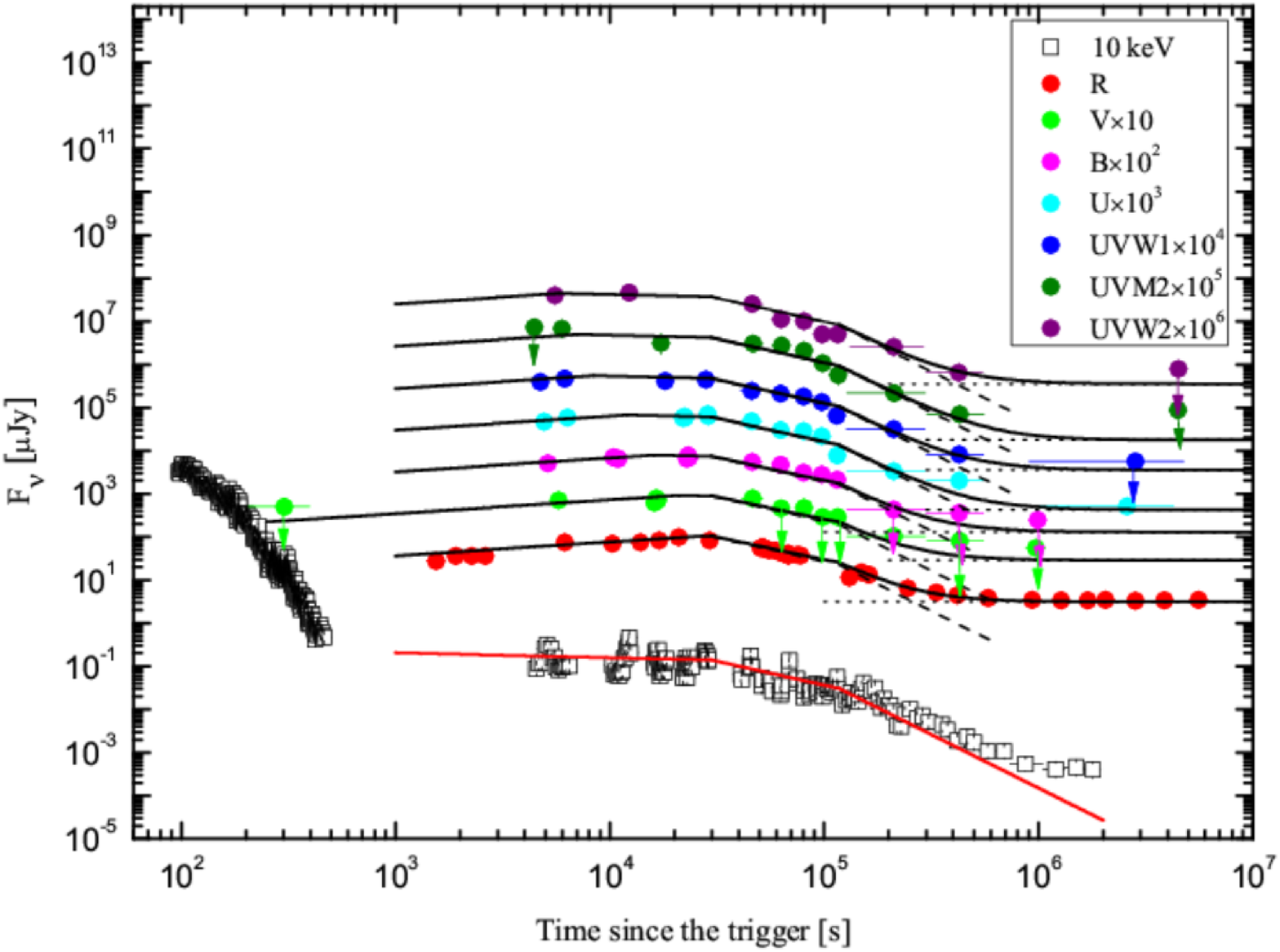}
\caption{Theoretical LCs as compared with the multi-band afterglow observations of GRB~060614. The 10~keV unabsorbed X-ray data (empty squares) are downloaded from
http://www.swift.ac.uk/burst\_analyser/00214805/ \citep{Evans07,Evans09}. The $R$-band data (red filled circles) are taken from \citet{dell06}
and \citet{gal06}, while the optical/UV data in other bands are take from \citetalias{mang07}. The optical/UV data have been corrected for
Galactic and host galaxy extinction with $A_{V,{\rm G}}=0.07$ and $A_{V,\rm h}=0.05$, respectively. For clarity, the shown flux densities in
the $V$, $B$, $U$, $UVW1$, $UVM2$ and $UVW2$ bands have been rescaled by factors 10, $10^2$, $10^3$, $10^4$, $10^5$ and $10^6$, respectively.
The red solid line is our model predicted X-ray LC. The modeled optical/UV LCs are shown as the sum (black solid lines) of two components: the afterglow (black dashed lines) and the host (black dotted lines).
The magnitude values of the host in each band are taken from \citetalias{mang07}. To produce the theoretical LCs, the parameters of $p_1=1.6$,
$p_2=2.68$, $q=0.41$, $e=0.27$, $\epsilon_{\rm e,-1}=1$, $\epsilon_{\rm B,-2}=1.1\times10^{-2}$, $E_{\rm f,52}=2.25$, $n_0=0.01$, $\xi_4=1.2$,
$t_{\rm f}=29.7$~ks and $t_{\rm j}=117.2$~ks are used.
\label{LCs}}
\end{figure}

Based on Equations (\ref{fv1}), (\ref{fv2}), (\ref{fv_post}), (\ref{vm3})--(\ref{fv3}) and our obtained parameters, we can now compare our model with the multi-band afterglow
LCs. As shown in Figure \ref{LCs}, the whole optical/UV and X-ray (except the last few data points) LCs can be well described with our model\footnote{The initial steep decay of the X-ray LC before about 500~s is likely the prompt emission tail \citepalias{mang07} which is not a concern of our model.}. Especially in the optical/UV band, our model successfully explained the initial frequency-dependent decay feature and the corresponding spectral evolution. Besides the two achromatic breaks $t_{\rm b,1}$ and $t_{\rm b,2}$, there is an chromatic break $t_{\nu}$ in the
optical/UV LCs. It denotes the time that $\nu_{\rm b}$ crosses an observational frequency $\nu$. For $\nu=R, V, B, U, UVW1, UVM2, UVW2$, the corresponding breaks are $t_\nu=26.8, 21.1, 16.1, 12.0, 8.4, 7.0, 5.8$~ks. The optical/UV LCs show a plateau between $t_\nu$ and $t_{\rm b,1}$
with the same slope as the X-ray plateau; before $t_\nu$, the LCs rise with a slope of ($-$0.32). It should be noted that an exact calculation
of afterglow radiation would give smooth spectral and temporal breaks \citep{Granot02}, so such a chromatic break  in the optical/UV LCs may not be clearly seen, especially when the data are sparsely sampled. Instead, the passage of $\nu_{\rm b}$ through the optical/UV band may show an average effect in the LCs: slowly rising at low frequencies and flattening at higher energies, just like the afterglow of GRB~060614 \citepalias{mang07}. We emphasize, however, that our simple analytic model perfectly described this feature and no need to employ complicated numerical calculations. For the X-ray afterglow, we note that the data points after $\sim10^6$~s obviously deviate from our modeling fit and suggest a late re-brightening or a flattening. \citetalias{mang07} found that
 at the end the observations have small signal to noise ratios and approach the XRT sensitivity limit. We thus do not consider this inconsistency. There are also slight excesses between $2\times10^5$ and $10^6$~s, this is because in our modeling we used the central value of 2.48 for the post-jet-break  slope. When the uncertainty of this parameter is considered, this problem would be alleviated.

We conclude this subsection by comparing the $q$ value of GRB~060614 with that of GRB~091127. For GRB~091127,
$q=0.64\pm0.08$ \citepalias{Zhang15}, while GRB~060614 gives $q=0.41\pm0.20$. These results imply a similar evolution behavior of the injection break, with $q\sim0.5$. However, at this stage it is premature to say that this represents a universal law of the injection break and more such events are needed to test this conjecture.
We emphasize that the values of $q$ for both bursts are reliable, since the consistency checks have been performed with various afterglow observational constraints.
Finally, we emphasize that this DPLH spectrum predicts an injection break frequency evolving as $\nu_{\rm b}\propto t^{-3\left(1+q\right)/4}=t^{\sim\left(-1.1\right)}$ (for $e=1$), which is substantially faster than $\nu_{\rm c}$ in a single PL hard electron spectrum model. Therefore, when this kind of spectral break along with flat spectra in the optical band is observed in afterglows, it provides strong support to the above conjecture.

\subsection{GRB~060908} \label{fitting2}
The parameters to be constrained are $p_1$, $p_2$, $q$, $\epsilon_{\rm{e}}$, $\epsilon_{\rm{B}}$, $\xi$, $E$ and $A_{\ast}$.
The observed spectral indices require $\nu_{\rm m}<\nu_{\rm optNIR}<\min{(\nu_{\rm b}, \nu_{\rm c})}<\nu_{\rm X}$, then we have
$p_1=2\beta_{\rm optNIR}+1=1.66^{+0.50}_{-0.58}$ and $p_2=2\beta_{\rm X}=2.34^{+0.50}_{-0.44}$.
The value of $q$ can in principle be determined by the observed X-ray decay index $\alpha_{\rm X}$. According to Equation (\ref{fv3}),
we get $q=4\left(\alpha_{\rm X}-p_2/2\right)/\left(p_2-2\right)=-0.6^{+9.0}_{-5.8}$. It is not strange that $q$ is badly constrained, since
$\alpha_{\rm X}$ has a very weak dependence on $q$ and it is mainly determined by $p_2$ which has large uncertainties.
Based on the results of Subsection \ref{fitting1}, we assume $q=0.5$ for GRB~060809 and test whether it is consistent with other observational
properties. With this value of $q$ and $\alpha_{\rm X}=1.12$, we obtain $p_2=2.2$. Given that $p_1$ and $p_2$ obtained from the spectral indices have relatively large uncertainties, for simplicity we adopt $p_1=1.6$ and $p_2=2.2$ in the following calculations.

To calculate the flux density $F_{\nu}$, we should first determine the order between $\nu_{\rm b}$ and $\nu_{\rm c}$. Below we give some arguments:
(i) the spectral analysis of \citet{Covino10} requires $\nu_{\rm b}(8000~{\rm s})\gtrsim 5.5\times10^{14}$~Hz, with $\nu_{\rm b}\propto t^{-1.25}$ we
have $\nu_{\rm b}(80~{\rm s})\gtrsim 0.7$~keV; (ii) at the end of the X-ray observations, $\nu_{\rm c}$ should not have crossed the X-ray band.
We simply require $\nu_{\rm c}(5\times10^5~{\rm s})\lesssim 1$~keV, with $\nu_{\rm c}\propto t^{1/2}$ we get $\nu_{\rm c}(80~{\rm s})\lesssim 3.0\times10^{15}$~Hz.
That is, at the beginning  of the observations ($\sim80$~s), $\nu_{\rm b}$ should be near the low-end of the XRT band, while $\nu_{\rm c}$ should be
near the high-end of the ultraviolet band, i.e., $\nu_{\rm m}<\nu_{\rm optNIR}<\nu_{\rm c}<\nu_{\rm b}<\nu_{\rm X}$.
As $\nu_{\rm b}$ decreases and $\nu_{\rm c}$ increases, the spectrum transits to $\nu_{\rm m}<\nu_{\rm optNIR}<\nu_{\rm b}<\nu_{\rm c}<\nu_{\rm X}$
and  eventually becomes $\nu_{\rm m}<\nu_{\rm b}<\nu_{\rm optNIR}<\nu_{\rm c}<\nu_{\rm X}$.

Since we have $\nu_{\rm X}>\nu_{\rm c}$ throughout the observations, the corresponding electrons may suffer from significant inverse Compton losses,
especially when $\epsilon_{\rm B}$ has very small values. When the synchrotron self-Compton (SSC) effect is considered, the cooling
frequency $\nu_{\rm c}$ would be reduced by a factor of $(1+Y)^{-2}$ and the X-ray flux would be suppressed by $(1+Y)^{-1}$, here $Y$ is the Compton parameter \citep{Sari01}. With the adopted parameters,  we derive the break frequencies
according to Equations (\ref{vm2w})--(\ref{vbw}) and  replace $\nu_{\rm c}$ with $\nu_{\rm c}(1+Y)^{-2}$, i.e.
\begin{eqnarray}\label{vm3w}
\nu_{\rm m}&=&3.4\times10^{7}\xi_4^{-1.33}\epsilon_{\rm e,-1}^{3.33}\epsilon_{\rm B,-4}^{1/2}E_{52}^{0.67}A_{\ast}^{-0.17}t_{\rm d}^{-1.67}{\rm Hz}, \\ \label{vb3w}
\nu_{\rm b}&=&7.8\times10^{13}\xi_4^2\epsilon_{\rm B,-4}^{1/2}E_{52}^{0.25} A_{\ast}^{0.25} t_{\rm d}^{-1.25}~{\rm Hz}, \\ \label{vc3w}
\nu_{\rm c}&=&3.9\times10^{16}\epsilon_{\rm B,-4}^{-3/2}E_{52}^{1/2}A_{\ast}^{-2}t_{\rm d}^{1/2}(1+Y)^{-2}~{\rm Hz}.
\end{eqnarray}

The Compton parameter can be estimated as follows \citepalias{Resmi08}.
For $\nu_{\rm m}\leqslant\nu_{\rm c}\leqslant \nu_{\rm b}$,
\begin{eqnarray}
Y &\approx& \frac{\nu_{\rm b}^{\rm IC} F_{\nu_{\rm b}}^{\rm IC}}{\nu_{\rm b} F_{\nu_{\rm b}}}  \nonumber \\
  &=& 2\gamma_{\rm b}\gamma_{\rm c} \zeta \left(\frac{\gamma_{\rm m}}{\gamma_{\rm b}}\right)^{p_1-1}  \nonumber \\
  &=& 670\epsilon_{\rm e,-1}\epsilon_{\rm B,-4}^{-1}(1+Y)^{-1}, \label{Y1}
\end{eqnarray}
where $\zeta\equiv F_{\nu,\rm max}^{\rm IC}/F_{\nu,\rm max}=n\sigma_{\rm T}r$, $F_{\nu_{\rm b}}$ and $F_{\nu_{\rm b}}^{\rm IC}$
are the synchrotron and SSC flux at $\nu_{\rm b}$, respectively, and  $\nu_{\rm b}^{\rm IC}\simeq 2\gamma_{\rm b}^2\nu_{\rm b}$.
We note that $Y$ is only a simply function of $\epsilon_{\rm e}$ and $\epsilon_{\rm B}$. As long as $\epsilon_{\rm B}\lesssim0.067\epsilon_{\rm e,-1}$, we have $Y\gtrsim 1$. Small values of $\epsilon_{\rm B}\lesssim0.01$ are required for GeV-detected bursts if the GeV emission arises from
external shocks \citep[e.g.,][]{kumar09,kumar10,beni15}, and are also supported by recent systematic studies using X-ray/optical \citep{sant14} or radio \citep{Duran14} afterglow observations. Such small values of $\epsilon_{\rm B}$ imply a large $Y$ and significant SSC losses,
which have important effects on the derived blastwave energy and thus the prompt emission efficiency \citep{beni15,beni16}. With this consideration,
Equation (\ref{Y1}) can be written as
\begin{equation}
Y\approx 25.9~\epsilon_{\rm e,-1}^{1/2}\epsilon_{\rm B,-4}^{-1/2}. \label{Y2}
\end{equation}
For $\nu_{\rm m}\leqslant\nu_{\rm b}\leqslant\nu_{\rm c}$,
\begin{eqnarray}
Y &\approx& \frac{\nu_{\rm c}^{\rm IC} F_{\nu_{\rm c}}^{\rm IC}}{\nu_{\rm c} F_{\nu_{\rm c}}}  \nonumber \\
  &=& 2\gamma_{\rm c}^2\zeta \left(\frac{\gamma_{\rm m}}{\gamma_{\rm b}}\right)^{p_1-1}\left(\frac{\gamma_{\rm b}}{\gamma_{\rm c}}\right)^{p_2-1} \nonumber \\
  & \propto& \zeta \gamma_{\rm m}^{0.6}\gamma_{\rm b}^{0.6}\gamma_{\rm c}^{0.8}(1+Y)^{-0.8}, \label{Y3}
\end{eqnarray}
where $\gamma_{\rm c}^{\rm IC}\simeq2\gamma_{\rm c}^2\nu_{\rm c}$, and $F_{\nu_{\rm c}}$ and $F_{\nu_{\rm c}}^{\rm IC}$
are the synchrotron and SSC flux at $\nu_{\rm c}$, respectively. After a simple derivation, Equation (\ref{Y3}) gives $Y\propto t^{-0.1}$.

The transition occurs at $t=t_{\rm bc}$, which can be obtained by solving $\nu_{\rm b}(t_{\rm bc})=\nu_{\rm c}(t_{\rm bc})$. We note that the value
of $Y$ is basically a constant throughout the observations; including the volution effect would only flatten the X-ray LC by $t^{0.1}$ after $t_{\rm bc}$. Given that our calculation of $Y$ is not sufficiently accurate, we do not consider its evolution and simply use Equation (\ref{Y2}) in the
following parameter estimations. This simplification is also consistent with the fitting results of a constant PL decay of the X-ray LC \citep{Covino10}.

We define the time at which $\nu_{\rm b}$ crosses a specified optical/NIR frequency as $t_{\rm b, \nu}$. For $t<t_{\rm b, \nu}$, we have
$\nu_{\rm m}<\nu<\min{(\nu_{\rm b}, \nu_{\rm c})}<\nu_{\rm X}$, then the optical/NIR flux density can be obtained from Equation (\ref{fv1w}),
\begin{equation}
F_{\nu}=1.75 \xi_4^{-0.4}\epsilon_{\rm e,-1} \epsilon_{\rm B, -4}^{0.65}E_{52}^{0.7}A_{\ast}^{0.95}t_{\rm d}^{-1.0}
\left(\frac{\nu}{\nu_{\rm R}}\right)^{-0.3}\mu{\rm Jy}, \label{FRw}
\end{equation}
where $\nu_{\rm R}$ is $R$-band frequency. For $t\geqslant t_{\rm b,\nu}$, we have $\nu_{\rm m}<\nu_{\rm b}<\nu<\nu_{\rm c}<\nu_{\rm X}$ and $F_{\nu}=F_{\nu}(t_{\rm b,\nu})\left(t/t_{\rm b,\nu}\right)^{-1.37}$ according to Equation (\ref{fv2w}).

The 10 keV flux can be derived from Equation (\ref{fv3w}), (\ref{vc3w}) and (\ref{Y2}),
\begin{equation}
F_{10{\rm keV}}=2.88\times10^{-5}\xi_{4}^{0.2}\epsilon_{\rm e,-1}^{1/2}\epsilon_{\rm B,-4}^{0.55}E_{52}^{1.02}A_{\ast}^{0.025}t_{\rm d}^{-1.12}\mu{\rm Jy}.
\label{Fxw}
\end{equation}

To obtain the remaining parameters, we use the following observational constraints: (i) $F_{\nu_{\rm R}}(500~{\rm s})=622~\mu{\rm Jy}$\footnote{This value has been corrected with the same Galactic and host galaxy extinction $E(B-V)=0.03$ \citep{Covino10}.}; (ii)
$F_{10\rm keV}(15.7~{\rm ks})=0.016~\mu{\rm Jy}$; (iii) $\nu_{\rm b}(80~{\rm s})\lesssim 1$~keV and $\nu_{\rm b}(8000~{\rm s})\gtrsim5.5\times10^{14}$~Hz. Using Equations (\ref{FRw}) and (\ref{Fxw}), conditions (i) and (ii) give
\begin{eqnarray} \label{cond1w}
\epsilon_{\rm B,-4}^{1.75} A_{\ast} &=& 13953.0 \epsilon_{\rm e,-1}^{-2} E_{52}^{-2.74}, \\ \label{cond2w}
\xi_{4}&=&3.78\times10^9\epsilon_{\rm e,-1}^{-2.5} E_{52}^{-5.1}\epsilon_{\rm B,-4}^{-2.75} A_{\ast}^{-0.125}.
\end{eqnarray}
Using Equation (\ref{vb3w}), condition (iii) gives
\begin{equation}
0.36\lesssim  \xi_{4}^2\epsilon_{\rm B,-4}^{1/2} A_{\ast}^{0.25} E_{52}^{0.25}\lesssim 0.5. \label{cond3w}
\end{equation}
Interestingly, Equation (\ref{cond3w}) leads to very  strict limits and we simply take
\begin{equation}
\xi_{4}^2\epsilon_{\rm B,-4}^{1/2} A_{\ast}^{0.25} E_{52}^{0.25}= 0.45. \label{cond4w}
\end{equation}

Like the case of GRB~060614, here we also adopt $\epsilon_{\rm e,-1}=1$.
We adopt a typical prompt emission efficiency of $\eta_{\gamma}=E_{\gamma,\rm iso}/(E_{\gamma,\rm iso}+E)=15\%$ \citep{beni16} for GRB~060908, this corresponds to $E_{52}=35$.
By substituting these values in Equations (\ref{cond1w}), (\ref{cond2w}) and (\ref{cond4w}),
we get $\epsilon_{\rm B,-4}=6.7$, $A_{\ast}=0.03$ and $\xi_{4}=0.42$. We note the value of $\epsilon_{\rm B}$ is well consistent with the statistical results of \citet{sant14} and \citet{Duran14}. It is also consistent with the results of \citet{beni16}, who re-analyzed the
prompt emission efficiency using X-ray afterglows and taken into account the SSC effect.

It is important to perform consistency checks of our model with the above parameters. According to Equations (\ref{vm3w}) and (\ref{vc3w}),
we have (i) $\nu_{\rm c}(80~\rm s)=4.5\times10^{15}$~Hz and
$\nu_{\rm c}(5\times10^5~{\rm s})=1.5$~keV; (ii) $\nu_{\rm m}(100~\rm s)=4.4\times10^{14}$~Hz. That is, at the very beginning $\nu_{\rm m}\lesssim\nu_{\rm opt}<\nu_{\rm c}$ is satisfied, and at the end $\nu_{\rm c}$ is at the low-end of the XRT band. We note that at 100~s, $\nu_{\rm m}$ is the same as the $R$-band frequency and this time should correspond to the peak of the LC. This seems to be inconsistent with the observations. However, as discussed below,
the data points before $\sim100$~s may be dominated by emission from the reverse shock.
Therefore, our obtained parameters are fully compatible with the observations and our afterglow modeling of GRB~060908 is self-consistent.

\begin{figure}[ht!]
\plotone{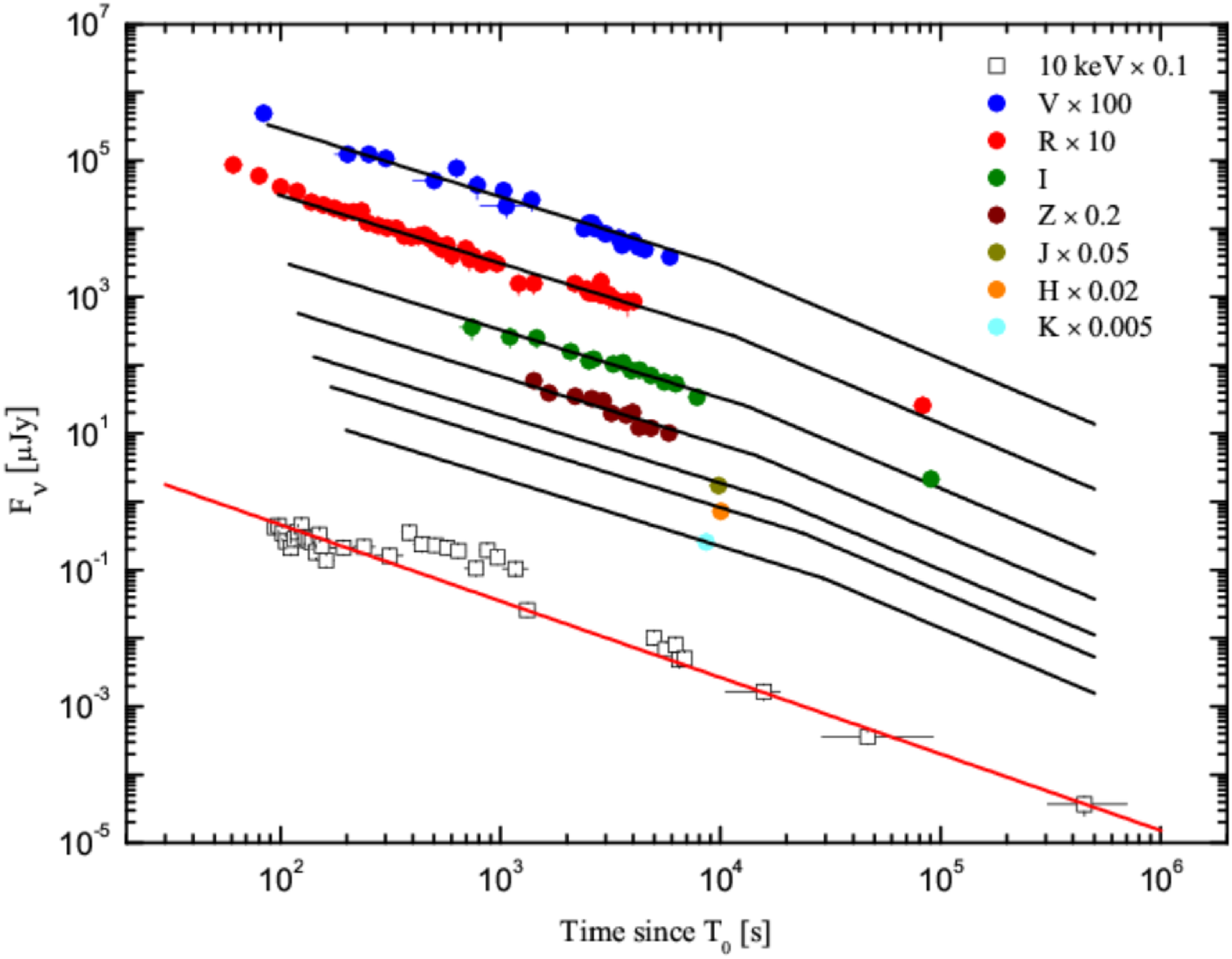}
\caption{Theoretical LCs as compared with the multi-band afterglow observations of GRB~060908. The 10~keV unabsorbed X-ray data (empty squares) are downloaded from
http://www.swift.ac.uk/burst\_analyser/00228581/ \citep{Evans07,Evans09}. The optical/NIR data (filled circles) are taken from \citet{Covino10}.
The optical/NIR data have been corrected with the same
Galactic and host galaxy extinction $E(B-V)=0.03$. For clarity, the shown flux densities have been rescaled by factors
ranging from 0.005 to 100.
The red and black solid lines are our afterglow modeling for the X-ray and the optical/NIR data, respectively.
To produce the theoretical LCs, the parameters of $p_1=1.6$,
$p_2=2.2$, $q=0.5$, $\epsilon_{\rm e,-1}=1$, $\epsilon_{\rm B,-4}=6.7$, $E_{52}=35$, $A_{\ast}=0.03$ and $\xi_4=0.42$ are used.
\label{LCs2}}
\end{figure}

Based on Equations (\ref{fv2w}), (\ref{vb3w}), (\ref{FRw}), (\ref{Fxw}) and the obtained parameters, we can now compare our model with the multi-band afterglow data. As shown in Figure \ref{LCs2}, the DPLH model can describe the afterglow rather well. The X-ray excesses between about 300 and 1000~s are likely due to a complex flaring activity and can be modeled with two Gaussian functions \citep{Covino10}. A more detailed analysis of flaring activity in this and other events was performed by \citet{chin10}. The predicted optical/NIR LCs initially decay as $t^{-1}$, then steepen to $t^{-1.37}$ at the frequency-dependent break time $t_{\rm b,\nu}$, which ranges from 9.8 to 29.3~ks for the observed bands. These values
are basically consistent with the fitting results ($\sim10^3-10^4$~s) of \citet{Covino10}. Unfortunately, observationally this break time cannot be well constrained by the data, let alone its chromaticity predicted by our model. The starting points of the black solid lines denote the times at which $\nu_{\rm m}$ crosses the corresponding bands, ranging from 88~s to 199~s, which could slightly change for different parameters adopted in the afterglow modeling.
The R-band data before $\sim100$~s exhibit an obvious excess component which decays as $t^{\sim\left(-1.4\right)}$. The origin of this component is not clear. One possibility is that the early decay is a superposition
of the decay phase \citep[$F_{\nu}\propto t^{-3}$][]{koba03} of the reverse shock emission and the smooth peak of the forward shock emission. This scenario is also compitable with our afterglow modeling which concerns only the forward shock emission.
Since the reverse shock component is not
distinctly identified in the LC, the relevant physical parameters cannot be constrained. We thus do not consider this physical process in our afterglow modeling.

\section{CONCLUSION AND DISCUSSION} \label{conclu}
The evidence for a DPLH spectrum in GRB afterglows have remained rare and somewhat ambiguous.
\citetalias{Zhang15} showed that GRB~091127
gave strong evidence for the existence of a DPLH spectrum with an injection break assumed as $\gamma_{\rm b}\propto \gamma^{q}$
in the highly relativistic regime. In this work, we show that GRB~060614 and GRB~060908 provide further evidence for such a spectrum.
We model the multi-band afterglow of GRB~060614 with the DPLH model in an ISM medium by taking into account a continuous energy injection process,
while for GRB~060908, a wind-like circumburst density profile is employed. The evidence for a DPLH spectrum is strong in the case of GRB~060614  since we directly see a spectral break passing through the optical/UV band, while in the case of GRB060908 the evidence appears to be less strong.
Perhaps most importantly, these bursts suggest a similar behavior in the evolution of the injection break, with $q\sim0.5$. Whether this represents
a universal law of the injection break remains uncertain and more such afterglow observations are needed to test this conjecture. 

Below we give some discussions on the afterglow modeling of GRB~060614:

Firstly, we mention the work of \citet{xu09} who also modeled the multi-band afterglow of this burst. Different from our model, they used a standard ($p>2$) electron energy spectrum and interpreted the observed spectral break as $\nu_{\rm m}$. This model was motivated by their SED analysis results at around 16~ks: the SED from the optical to X-ray bands was fitted by a broken PL which gave $\beta_{\rm opt}=-0.1\pm0.4$ (90\% CL)and $\beta_{\rm X}=0.9\pm0.1$ (90\% CL). This spectrum is compatible with $\nu_{\rm opt}<\nu_{\rm m}<\nu_{\rm X}<\nu_{\rm c}$ in the standard afterglow model. When an energy injection is assumed, this model can describe the afterglow LCs rather well \citep{xu09}. However, we should note that the value of $\beta_{\rm opt}$ cannot be well constrained in their fitting. Alternatively, the authors also fitted this SED by fixing $\beta_{\rm X}=\beta_{\rm opt}+0.5$ and gave $\beta_{\rm opt}\simeq0.36$ and $\beta_{\rm X}\simeq0.86$. These values are remarkably consistent with those of \citetalias{mang07}. Both works actually favor a positive $\beta_{\rm opt}$ which is not compatible with $\nu_{\rm opt}<\nu_{\rm m}$. Therefore,
their model has difficulties in explaining the early flat spectra in the optical/UV band.

Secondly, our model assumes an additional energy injection process, we now discuss its possible origins. The energy injection can be provided by
the central engine, e.g., a rapidly spinning, strongly magnetized
neutron star \citep[the so-called ``millisecond magnetar''; e.g.,][]{Usov92,Thom94,Dai98,Zhang01,Zhang06,rowl13,gomp14}.
However, the simplest dipole spin-down model predicts
$e=0$ that is not consistent with our obtained $e\sim 0.27$. Modifications to the simplest model are needed for this burst.  Alternatively,
such an energy injection can be provided by the soft tail of the outflow by considering that the extended emission is
several times more energetic than
the initial hard pulse. To check the consistency with our obtained parameters in Subsection \ref{fitting1}, we assume the energy injection takes place at $t_{\rm i}=100$~s and take $t_{\rm f}=29.7$~ks,
$E_{\rm f}=2.25\times10^{52}$~erg,  then the initial kinetic energy of the outflow is $E_{\rm i}=E_{\rm f}\left(t_{\rm i}/t_{\rm f}\right)^{\left(1-e\right)}\sim3.5\times10^{50}$~erg. Since the isotropic energy of the initial pulse is also $\sim3.5\times10^{50}$~erg \citepalias{mang07}, this corresponds
a radiation efficiency of $\sim50\%$, which is consistent the median value of short GRBs \citep{fong15}.

Finally, the steep post-jet-break decay of GRB~060614 is due to significant sideways expansion of the jet based on the theory of \citet{Rhoads99}. However, our derived bulk Lorentz factor at the jet-break time is $\sim 11$ that is highly relativistic. This is in conflict with the results given by numerical simulations and more sophisticated analytical treatments which suggest that the sideways expansion of a relativistic jet is not important until $\gamma$ drops below $\sim 2$ \citep{Huang00,Granot01,Kumar03,Cann04,ZhangW09,decolle12,Granot12,van12}.
There are also numerical works \citep[e.g.,][]{wygoda11} supporting the simple analytic solutions of \citet{Rhoads99}.  Nevertheless, \citet{Granot12} found that exponential sideways expansion can only occur for jets with extremely narrow initial half-opening angle ($\theta_0\ll0.05$) when $\gamma\lesssim 1/\theta_{\rm j}$ is satisfied. Considering that the realistic GRB jets may have much more complicated hydrodynamical
evolutions than employed in the above analytic and numerical models, whether an early exponential sideways expansion phase exists for typical
jet opening angles still remains uncertain. Observationally, a fraction of X-ray afterglows show a jet-break-like feature at around 1 day with post-break slope of $\sim p$ \citep[e.g.,][]{Zhang06,willi07,Liang08,Evans09,Racu09,panai12}; some recent systematic studies of multi-band afterglows have also
shown that a small fraction of GRBs have such jet-break features, simultaneously in X-rays and in the optical band \citep[e.g.,][]{fong15,Li15,Wang15}. Therefore, at least for some GRBs, sideways expansion should be significant when $\gamma\lesssim 1/\theta_{\rm j}$ is satisfied, even though the jet is still in the highly relativistic regime.

Unlike GRB~060614 and GRB~091127 that clearly show both spectral breaks passing through the optical bands and chromatic evolutions in the multi-band LCs, the afterglow of GRB~060908 exhibits no such features. Such spectral breaks were interpreted as the injection break frequency ($\nu_{\rm b}$) and the corresponding chromatic breaks were due to the passage of $\nu_{\rm b}$ through the optical bands.
For GRB~060908, our model predicts that $\nu_{\rm b}$ crosses the observed optical bands between 9.8 and 29.3~ks, at which the chromatic breaks should be seen. Unfortunately, the data around this time are not sufficient to perform detailed spectral and temporal analysis. Nevertheless, the fact that the optical LCs show marginal evidence for a break at around $10^3-10^4$~s with a consistent post-break slope with our model prediction provides an additional support to the DPLH model.

Last but not least, we want to give a discussion on the value of $q$. In the original work of \citet{B04}, the authors assumed the injection break to be the minimum electron Lorentz factor that can be accelerated by relativistic shocks, i.e., $\gamma_{\rm b}\equiv\gamma_{\rm acc}=\left(m_{\rm p}/m_{\rm e}\right)\gamma$; between $\gamma_{\rm m}$ and $\gamma_{\rm b}$, some other acceleration mechanisms take place and produce a hard electron spectrum. This means $q=1$ in their model. However, our work suggests $q\sim0.5$ that disfavors this scenario. Moreover, the
value of $\xi$ we derived is much larger than $m_{\rm p}/m_{\rm e}$.
\citetalias{Resmi08} extended this function by assuming $\gamma_{\rm b}\propto (\beta\gamma)^q$ and attempted to find evidence by modeling the afterglows
of three pre-{\it Swift} GRBs. Although their model can explain the afterglow LCs, the evidence for a DPLH spectrum is far from robust. First, no
bursts in their sample show very flat spectra in the optical band. Their spectral indices are in the range of 0.6--0.9 that is typical for optical afterglows \citep[e.g.,][]{Li12}. Second, no spectral evolution was seen in their sample. That is, the injection break frequency $\nu_{\rm b}$ was
actually not observed directly. Finally, for these bursts, the DPLH model is not the sole explanation. The LCs can also be reproduced by a model assuming continuous energy injection \citep[e.g.,][]{Bjor02}. Besides, their derived model parameters are much different from ours (see Paper I for
a detailed discussion). Especially on the value of $q$, they gave $q\gtrsim 1$ for all bursts, while ours is substantially smaller. Since our works have provided the most robust evidence for a DPLH spectrum so far, $q\sim0.5$ should be preferred.

The origin of the hard electron energy distribution is not clear. Our results may offer guidance in the right direction.
Meanwhile, more observations of GRB afterglows with a hard electron spectrum and further developments in the area of simulations of the {\it Fermi} acceleration process in relativistic shocks will help us understand the origin of the observed spectra of GRBs and their afterglows.

\acknowledgments
We acknowledge the anonymous referee for helpful comments and suggestions.
This work made use of data supplied by the UK Swift Science Data Centre at the University of Leicester.
This study was supported by the Strategic Priority Research Program of the Chinese Academy of Sciences (Grant No. XDB23040400).
SLX was also supported by the Hundred Talents Program of the Chinese Academy of Sciences (Grant No. Y629113).
LMS acknowledges support from the National Program on Key Research and Development Project (Grant No. 2016YFA0400801)
and the National Basic Research Program of China (Grant No. 2014CB845802).

\end{document}